\documentclass[12pt]{iopart}
\usepackage{times}
\usepackage{graphicx}

\begin{document}

\title{One-parameter nonrelativistic supersymmetry for microtubules}
\author{H C Rosu\footnote{e-mail: hcr@ipicyt.edu.mx},  J M Mor\'an-Mirabal and O Cornejo}
\address{Instituto Potosino de Investigaci\'on Cient\'{\i}fica y Tecnol\'ogica, Apdo Postal 3-74 Tangamanga, San Luis Potos\'{\i}, MEXICO}
\begin{abstract}
The one-parameter nonrelativistic supersymmetry of Mielnik  [J. Math. Phys. 25, 3387 (1984)] is applied to the 
simple supersymmetric model of Caticha [Phys. Rev. A 51, 4264 (1995)] in the form used by Rosu [Phys. Rev. E 55, 2038 (1997)]
for microtubules.
By this means, we introduce Montroll double-well potentials with singularities that move along the positive or negative traveling direction
depending on the sign of the free parameter of Mielnik's method. Possible interpretations of the singularity are either microtubule
associated proteins (motors) or structural discontinuities in the arrangement of the tubulin molecules. 
\end{abstract}

{

PACS number(s):
87.15.He, 03.65.Ge, 11.30.Pb 

\bigskip




Microtubules (MTs) are hollow cylinder tubes, 25 nm in outer diameter and 14 nm inner diameter, made of two types of 4 nm-long dimers of a polar protein
known as tubulin that can self-assemble both {\em in vivo} and {\em in vitro} to lengths from several nm up to mm in some neurons. 
They form the main filamentary component of the cytoskeleton of
all eukaryotic cells. Along their walls the tubulin dimers are distributed onto 13 (the seventh Fibonacci number) so-called protofilaments laterally associated.
Brain tissues are especially enriched in MTs. Many interesting speculations on MTs have been advanced in recent years \cite{spec}. 

Based on well-established results of Collins, Blumen, Currie and Ross \cite{1} regarding the dynamics of domain walls in ferrodistortive 
materials, 
Tuszy\'nski  and collaborators \cite{mtub1,mtub2} considered MTs to be ferrodistortive and studied 
kinks of the Montroll type \cite{mont} as excitations responsible for the
energy transfer within this highly interesting biological context.

The Euler-Lagrange dimensionless equation of motion of ferrodistortive domain walls as derived in \cite{1}
from a Ginzburg-Landau free energy with driven field and dissipation included is of the traveling reaction-diffusion type
\begin{equation} \label{1}
\psi ^{''}+\rho\psi ^{'}-\psi ^3 +\psi+\sigma=0~,
\end{equation}
where the primes are derivatives with respect to a traveling coordinate
$\xi =x-vt$, $\rho$ is a friction coefficient and $\sigma$ is related to
the driven field \cite{1}.

There may be ferrodistortive domain walls that can be identified with the Montroll  kink solution of Eq.~(\ref{1})
\begin{equation}  \label{2}
M(\xi)=\alpha _1+\frac{\sqrt{2}\beta}{1+\exp(\beta\xi)}~,
\end{equation}
where $\beta=(\alpha _2-\alpha _1)/\sqrt{2}$
and the parameters $\alpha _1$ and $\alpha _2$ are two nonequal solutions of the cubic equation
\begin{equation} \label{3}
(\psi -\alpha _1)(\psi -\alpha _2)(\psi -\alpha _3)=\psi ^3 -\psi -\sigma~.
\end{equation}

In a previous paper \cite{rosu}, one of the authors noted that
Montroll's kink can be written as a typical $\tanh$ kink
\begin{equation} \label{M}
M(\xi)=
\gamma -\tanh\left(\frac{\beta \xi}{2}\right)~,
\end{equation}
where $\gamma \equiv \alpha _1 +\alpha _2=1+\frac{\alpha _1\sqrt{2}}{\beta}$.
The latter relationship allows one to use a simple construction method of exactly
soluble
double-well potentials in the Schr\"odinger equation proposed by Caticha
\cite{cat}.
The scheme is a non-standard application of
Witten's supersymmetric quantum mechanics \cite{w}
having as the essential assumption the idea of considering the traveling $M$ kink as
the switching function between the two lowest eigenstates of the
Schr\"odinger equation with a double-well potential. Thus
\begin{equation} \label{phiM}
\phi _1=M\phi _0~,
\end{equation}
where $\phi _{0,1}$ are solutions of $\phi ^{''}_{0,1}+[\epsilon _{0,1}-u(\xi)]
\phi _{0,1}(\xi)=0$, and $u(\xi)$ is the double-well potential to be found.
Substituting Eq.~(\ref{phiM}) into the Schr\"odinger equation for
the subscript 1 and substracting the same equation multiplied by the
switching function for the subscript 0, one obtains
\begin{equation} \label{phiR}
\phi ^{'}_{0}+R_M\phi _0=0~,
\end{equation}
where $R_M$ is given by
\begin{equation} \label{R}
R_M=\frac{M^{''}+\epsilon M}{2M^{'}}~,
\end{equation}
and $\epsilon=\epsilon _1-\epsilon _0$ is the lowest energy splitting in
the double-well Schr\"odinger equation. 
In addition, notice that Eq.~(\ref{phiR}) is the basic equation introducing the superpotential $R$ in
Witten's supersymmetric quantum mechanics, i.e., the Riccati solution.
For Montroll's kink the corresponding Riccati solution reads
\begin{equation} \label{RM}
R_M(\xi)=-\frac{\beta}{2}{\rm tanh}\left(\frac{\beta}{2}\xi\right)+\frac{\epsilon}{2\beta}\Bigg[\sinh(\beta\xi)+
2\gamma\cosh ^2\left(\frac{\beta}{2}\xi\right)\Bigg]
\end{equation}
and the ground-state Schr\"odinger function is found by means of Eq.~(\ref{phiR})
$$
\phi _{0,M}(\xi) =\phi _0(0)\cosh\left(\frac{\beta}{2}\xi\right)\exp\left(\frac{\epsilon}{2\beta ^2}\right) 
$$
\begin{equation} \label{phi0}       
\exp\left(-\frac{\epsilon}{2\beta ^2}\Big[
\cosh (\beta \xi)-\gamma \beta\xi -\gamma\sinh(\beta\xi)\Big]\right)~,
\end{equation}
while $\phi _1$ is obtained by switching the ground-state wave function by means of $M$. 
This ground-state wave function is of supersymmetric type
\begin{equation} \label{phi}
\phi _{0,M}(\xi)=\phi _{0,M}(0)\exp\Bigg[-\int_0^{\xi} R_M(y)dy\Bigg]~,
\end{equation}
where $\phi _{0,M}(0)$ is a normalization constant. 

The Montroll double well potential is determined up to the additive constant $\epsilon _0$ by the `bosonic' Riccati equation
$$
u_M(\xi)=R_M^2-R_M^{'}+\epsilon _0= 
\frac{\beta ^2}{4}+\frac{(\gamma ^2 -1)\epsilon ^2}{4\beta ^2}+\frac{\epsilon}{2}+\epsilon _0+
$$
$$
+\frac{\epsilon}{8\beta ^2}\Big[\left(4\gamma ^2\epsilon +(2\gamma ^2+1)\epsilon {\rm cosh} (\beta \xi )-8\beta ^2\right){\rm cosh} (\beta \xi )-
$$
\begin{equation} \label{u}
-4\gamma \left(\epsilon +\epsilon {\rm cosh}(\beta \xi)-2\beta ^2 \right) {\rm sinh } (\beta \xi)\Big]~.
\end{equation}
Plots of the asymmetric Montroll potential and ground state wavefunction are given in Figs.~(1) and (2) for a particular set of the parameters. 
If, as suggested by Caticha, one chooses the ground state energy to be
\begin{equation} \label{eps}
\epsilon _0=-\frac{\beta ^2}{4}-\frac{\epsilon}{2}+\frac{\epsilon ^2}{4\beta ^2}
\left(1-\gamma ^2\right)~,
\end{equation}
then $u_M(\xi)$ turns into a traveling, asymmetric Morse double-well potential of
depths depending on the Montroll parameters $\beta$ and $\gamma$ and the splitting $\epsilon$
\begin{equation} \label{U}
U_{0, m}^{L,R}=\beta ^2\Bigg[1\pm \frac{2\epsilon \gamma}{(2\beta)^2}\Bigg]~,
\end{equation}
where the subscript $m$ stands for Morse and the superscripts $L$ and $R$ for left and right well, respectively.
The difference in depth, the bias, is
$\Delta _m\equiv U_0^L-U_0^R=2\epsilon\gamma$, while the location of the
potential minima on the traveling axis is at
\begin{equation}  \label{xiLR}
\xi _{m}^{L,R}=\mp\frac{1}{\beta}\ln \Bigg[\frac{(2\beta)^2\pm
2\epsilon\gamma}{\epsilon(\gamma\mp 1)}\Bigg]~,
\end{equation}
that shows that $\gamma \neq \pm 1$.

A one-parameter supersymmetric extension of the previous results is possible. It is quite known in the 
literature on supersymmetric quantum mechanics where it has been introduced by Mielnik, Fernandez and Nieto \cite{mielnik}
and is based on the Darboux covariant isospectrality of Schroedinger equations.  
In the biological context it has been applied to the DNA molecule by Drigo-Filho and Ruggiero \cite{dfr}. 
The point is that $R_M$ as given in 
Eq.~(\ref{RM}) is only the particular solution of the Riccati equation ocurring in Eq.~(\ref{u}). A more general, parameter-dependent 
Riccati equation of the form $u_M(\xi ; \lambda)=R_M^2(\xi ; \lambda)-R_M^{'}(\xi ;\lambda)+\epsilon _0$ can be constructed
whose solution is a one-parameter function of the form
\begin{equation} \label{genR}
R_M(\xi ;\lambda)=R_M(\xi)+\frac{d}{d\xi}\Big[\ln(I_M(\xi)+\lambda)\Big]
\end{equation} 
and the corresponding one-parameter Montroll potential is given by
\begin{equation} \label{genu}
u_M(\xi ;\lambda)=u_M(\xi)-2\frac{d^2}{d\xi ^2}\Big[\ln(I_M(\xi)+\lambda)\Big]~.
\end{equation} 
In these formulas, $I_M(\xi)=\int ^{\xi}\phi _{0,M}^2(\xi)d\xi$ and $\lambda$ is an integration constant
that is used as a deforming parameter of the potential and is related to the irregular zero mode. 
The one-parameter Darboux-deformed ground state wavefunction can be shown to be
\begin{equation} \label{wfl}
\phi_{0,M}(\xi ;\lambda)=\sqrt{\lambda(\lambda+1)}\frac{\phi _{0,M}}{I_M(\xi)+\lambda}~,
\end{equation} 
where $\sqrt{\lambda(\lambda+1)}$ is the normalization factor implying that $\lambda \notin [0,-1]$. Plots of $u_M(\xi ; \lambda)$ and
$\phi _{0,M}(\xi; \lambda)$ for $\lambda =10$ 
are presented in Figs.~(3) - (4). See also  Fig.~(5 ) for a plot of the function $I_M(\xi)$ producing the parametric Darboux deformation. 
A singularity at $I_M(\xi)+\lambda=0$ is 
introduced in both potential and wavefunction.  If the parameter $\lambda$ is positive the singularity is to be found on the negative $\xi$ axis, 
while for negative $\lambda$ it is on the positive side.
For large values of $\pm\lambda$ the singularity moves towards $\mp \infty$ and the potential and ground state wave 
function recover the shapes of the non-parametric potential and wavefunction. 
The one-parameter Morse case corresponds formally to the change of subscript $M\rightarrow m$
in Eqs.~(\ref{genR}) and (\ref{genu}).
For the single well Morse potential  the one-parameter procedure has been studied by
Filho \cite{filho} and Bentaiba {\em et al} \cite{bentaiba}.
Potentials and wavefunctions with singularities are not so strange as it seems \cite{cs}. Similar to the case of the $\delta$ potential in condensed
matter physics, we interpret the singularity
as representing the effect of an impurity moving along the microtubule in one direction or the other depending on the sign of the parameter
$\lambda$. In the case of microtubules, The impurity may represent a protein attached to the microtubule or a structural discontinuity in the 
arrangement of the tubulin molecules. This interpretation of impurities has been given by Trpi\v{s}ov\'a and Tuszy\'nski in 
non-supersymmetric models of nonlinear microtubule excitations \cite{tt}. 

In conclusion, the supersymmetric approaches allow for a number of interesting
{\em exact} results and point to a direct connection between Schroedinger double-well potentials and 
nonlinear kinks encountered in nonequilibrium chemical processes. MTs are an important application
but the procedures described here can be used in many other applications. Moreover, the supersymmetric constructions can 
be used as a background for clarifying further details of the exact models. Although it is not so clear why one should 
take a certain type of kink as switching function between the Schroedinger split modes, it is interesting that proceeding
in this way one will be led to some familiar double-well potential in chemical physics.



\begin{figure}
\begin{center}
\includegraphics{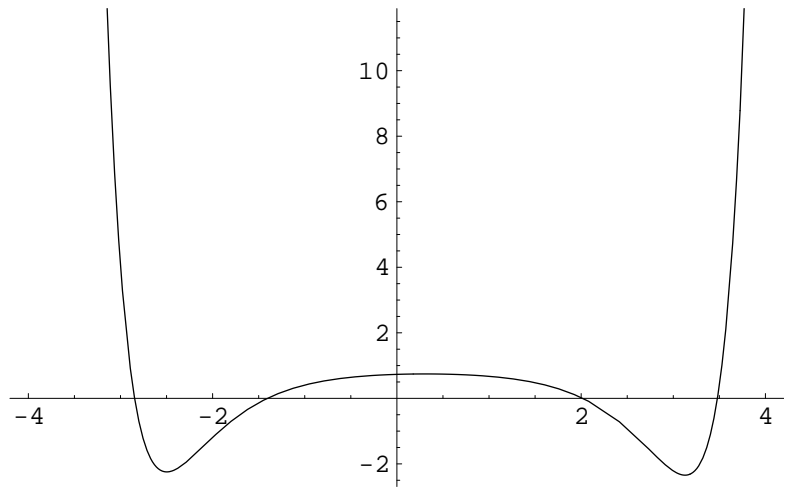}
\end{center}
\caption{The Montroll asymmetric double-well potential (MDWP) calculated using Eq.~(11) for 
$\epsilon _0=0$. In all figures $\alpha _1 =1$, $\alpha _2=-1.5$, $\beta =-2.5/\sqrt{2}$, $\gamma =-0.5$, 
$\epsilon =0.1$.}
\label{boring figure1}
\end{figure}

\begin{figure}
\begin{center}
\includegraphics{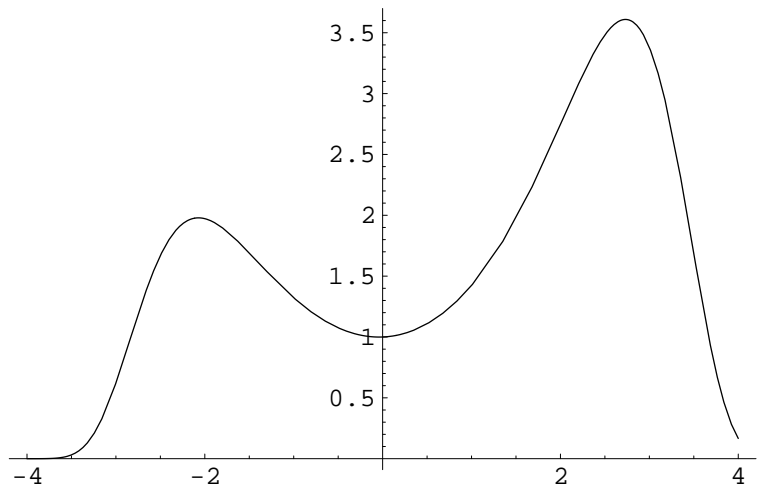}
\end{center}
\caption{The Montroll ground state wave function cf. Eq. ~(9) for $\phi _{0}(0)=1$.}
\label{boring figure2}
\end{figure}

\begin{figure}
\begin{center}
\includegraphics{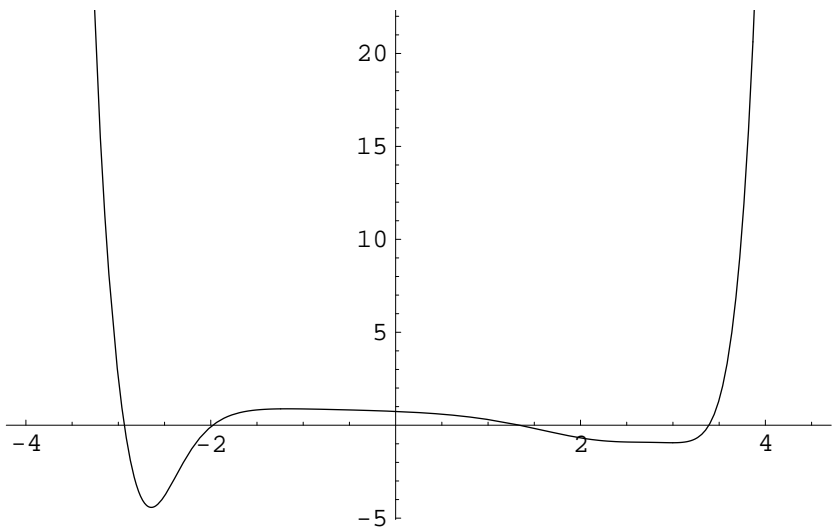}
\end{center}
\caption{The one parameter Darboux-modified MDWP for $\lambda =10$}
\label{boring figure3}
\end{figure}

\begin{figure}
\begin{center}
\includegraphics{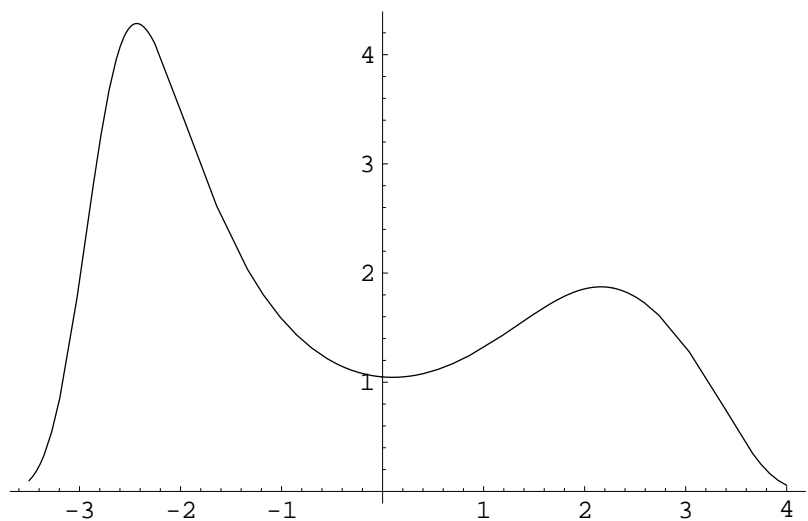}
\end{center}
\caption{The ground state wave function corresponding to $\lambda =10$.}
\label{boring figure4}
\end{figure}



\begin{figure}
\begin{center}
\includegraphics{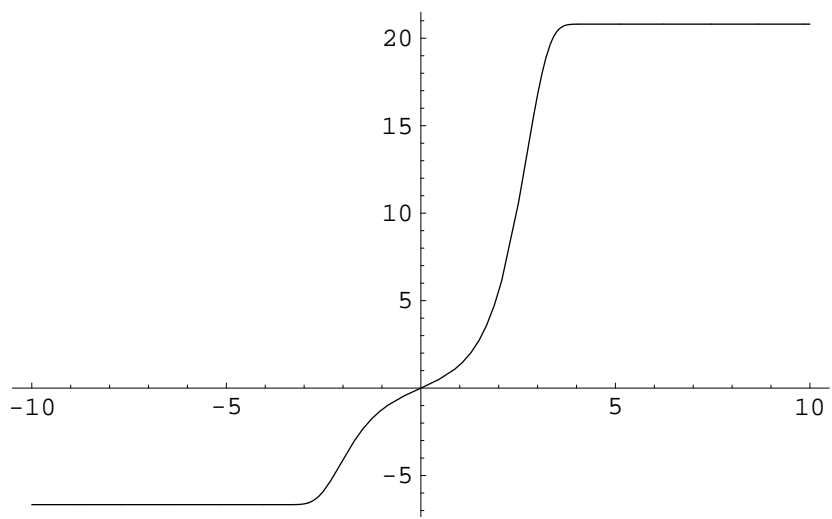}
\end{center}
\caption{Plot of the integral $I_M(\xi)$ that produces the deformation of the potential and wavefunctions.}
\label{boring figure2}
\end{figure}





\end{document}